\documentclass[12pt,letterpaper,onecolumn]{article}
\usepackage[latin1]{inputenc}
\usepackage{amsmath}
\usepackage{amsfonts}
\usepackage{amssymb}
\usepackage{braket}
\usepackage[left=1.00in, right=1.00in, top=1.00in, bottom=1.00in]{geometry}
\author{Horace P. Yuen\\
Department of Electrical Engineering and Computer Science\\
Department of Physics and Astronomy\\
Northwestern University, Evanston IL. 60208\\
yuen@eecs.northwestern.edu
}
\title{Problems of Security Proofs and Fundamental Limit on Key Generation Rate in Quantum Key Distribution}
\begin{document}
\maketitle
\begin{abstract}
\textit{It is pointed out that treatments of the error correcting code in current quantum key distribution protocols of the BB84 type are not correct under joint attack, and the general interpretation of the trace distance security criterion is also incorrect. With correct interpretation of the criterion as well as a correct treatment of the error correcting code and privacy amplification code, it is shown that even for an ideal system under just collective attack, the maximum tolerable quantum bit error rate is about 1.5$\%$ and a net key cannot actually be generated with practical error correcting codes even at such low rate, contrary to claims in the literature.}
\end{abstract}

\section{Introduction}
In quantum key distribution (QKD), quantum effect that has no classical analog is utilized to generate a sequence of bits, the secret key $K$, between two parties A and B that are known only between themselves. The typical approach involves information-disturbance tradeoff in BB84 type protocols [1], but other quantum approach without using such tradeoff is possible [2]. In the most recent and complete treatment [3] of the fundamental security and key rate of a single-photon BB84 protocol that incorporates some but not all unavoidable system imperfections, security is quantified by a trace distance criterion and the secure key rate is computed by subtracting from the generated key $K$ a term $leak_{EC}$ which accounts for information leak due to error correction to give the net key rate. The use of such criterion and $leak_{EC}$ is currently the widespread approach to QKD security analysis. However, unless the incorrect interpretation of the trace distance criterion [4,5] is used, the generated key is far from uniform [6,7] and cannot be used to account for the uniform shared secret key bits in $leak_{EC}$ that are used in such error correcting code treatment in the security proof. As a result, it is shown in this paper that privacy amplification has to work much harder quantitatively even if one accepts just near-uniform bits for $leak_{EC}$ subtraction and even if only collective attacks are considered. It will be explained why the key generation rate under joint attack has not been quantified and may not be quantifiable. When the correct operational meaning of the criterion is employed, it will be shown that the resulting tolerable quantum bit error rate (QBER) for net key generation under collective attacks without any imperfection in the scheme of [3] is $\sim$ 1.5\%, and that no known error correcting code (ECC) can be found that would actually produce a positive net key. If a multi-photon source such as a laser is employed, it would appear impossible to produce a secure net key even when the message authentication bits necessary for thwarting man-in-the-middle attack in such disturbance estimation protocols are not counted. This is especially the case, apart from detector electronics loopholes [8], when unavoidable system imperfections are taken into account. Thus, it appears new approaches are needed in QKD if it is ever going to practically produce a secure net key.

\section{Problem Formulation and Error Correction}
Let $S$ be the sifted key (raw key) of length $|S|$ after basis matching and disturbance checking with acceptable QBER given by $Q$ in a run of the BB84 protocol. Let $L$ be the key after error correction on $S$, and $K$ the generated key of length $|K|$ after a privacy amplification code (PAC) is applied to $L$. In the current standard security approach [3], the quantity $H^{\epsilon}_{min}$ is bounded from the observed $Q$. We dispense with the system imperfections in the protocol of [3] to concentrate just on the fundamental issues in connection with ECC and PAC, and will be mainly concerned with $H_{min}$, which is $H_{min}^\epsilon$ for $\epsilon \sim 2^{-|K|} \sim 0$. Let the lower case $s$ represent a specific value that the random variable $S$ takes for a probe state $\rho^{s}_E$ on the probe E sets during her attack on the sent quantum signals that led to $s$. The $H_{min}$ is simply related [9] to the optimal $S$-averaged probability $\bar{p}_1(S)$ that E estimates the whole $S$ correctly from $\rho^{s}_E$ and the a priori distribution of $S$ according to quantum detection theory [10]. From a lower bound on $H_{min}$ equivalent to an upper bound on $\bar{p}_1(S)$, $S$ is then taken as the input $L$ to the PAC in assessing the final security of $K$ according the trace distance $d$ criterion on $\rho^k_E$ [4,5]. No specific ECC is mentioned and the amount of bits given by [3,11]
\begin{equation}
leak_{EC}=f\cdot|S|\cdot h(Q)
\end{equation} where $h(\cdot)$ is the binary entropy function and $f$ an "efficiency" factor, is subtracted from $|K|$ to account for information leaking from error correction, to yield the bit length for the \textit{net} generated key. The value of $f$ is typically arbitrarily chosen to be $f$=1.1--1.2. The justification of (1) in [3] for ideal $f$=1 is given by citation of the whole book [12] that does not deal with such problem at all. One may surmise the actual justification as follows.

For two-way reconciliation to correct B's errors such as the Cascade protocol, there is no general quantification on how much information is leaked to E and how the system performs, in addition to errors of reasoning in Cascade itself [13]. That protocol is highly nonlinerarly random and resists quantitative analysis, as would any two-way interactive protocol. Padding all open parity bits may cost a prohibitively large number of shared secret bits. One would find the asymptotic limit (1) with $f$=1 if an agent knowing exactly where the errors occur for the $S$ in B's possession is telling B where they are, from the large $|S|$ behavior of the binomial coefficient. This is clearly not applicable to reconciliation where the error positions are unknown to the users. In the literature, it seems the only sketch of argument for (1) ever given is by equ (4)-(5) in ref [14] for one-way error correction in the ideal limit $f$=1.

The channel capacity between the two users under QBER=$e$ is taken to be $I_S(e)=1-h(e)$ per use, that of a memoryless binary symmetric channel [12]. The "minimum number of bits needed, on average, to correct a key of length n affteced by the error rate $e$ is the given by" [14]
\begin{equation}
n_{min} = n[1-I_S(e)].
\end{equation} Since (2) can be written as $n_{min}=n-k$ for $k=nI_S(e)$, the following argument for (2) may be derived from the discussion in section III.B.1 of [14]. A linear $(m,n)$ ECC [15] is used on $S$ with now $n=|S|$. A prior shared secret key is used to one-time pad the transmission of the $m-n$ parity-check bits [16] of the codeword with information bits $S$. This can always be done by writing the linear code in systematic form. If the padding is done with a bit string $U$ that is uniformly distributed to E, there will be no leak of information to E. Thus the corrected $S$ and $L$ can be considered identical as input to the PAC, since E should not know more about the corrected $S$ (from A) than what she knows on $S$ (for B).

This subtraction is not valid. Under joint attack by E the user's channel is not memoryless, which is the case even if collective attack (identical probe from E on each qubit) is optimum for E from the viewpoint of her information gain [17] because E may launch a joint attack to minimize the users' key rate. This is because the user's error probability depends on the underlying channel statistics, in particular the probability of various patters of errors in $S$. This is why the memoryless binary symmetric channel capacity is $1-h(Q)$, and not simple $1-2Q$ from correction on the number of errors asymptotically. This problem underlies \textit{all} existing joint-attack security proofs, in particular the classical counting argument [18] on estimating the QBER in $S$ from the checked qubits is \textit{not} valid. Even if a valid quantum estimate can be derived, it would surely produce much wider fluctuation than the classical counting result. This quantum fluctuation estimate issue persists even in an algebraic error correction approach in lieu of probabilistic decoding. A full treatment of these issues will be given elsewhere. 

Under collective attacks so that (2) applies, the number of added parity-checked bits to S is given by $\frac{|S|}{1-h(Q)} - |S|$ or \begin{equation}
leak_{EC}=\frac{h(Q)}{1-h(Q)}|S|
\end{equation}There is yet no guarantee one can find a polynomial decodable [19] linear ECC that would give the above code rate. For a concrete finite protocol, the real issue is whether an $(m,n)$ linear ECC can be found with a low enough resulting error probability for the users while practically decodable. Thus, we have shown that in a fundamental sense \textit{no} concrete protocol has yet been given with a provable net secure key rate.

It is important to observe that the other common approach on the ECC in QKD simply ignores E's ECC information and is thus not valid. This is because the security condition on $S$ cannot be otherwise transferred to $L$ in assessing  the security of $K$. This happened in the early proofs such as [20], and in the situation when memoryless channel model applies as in the proof of [21] and in classical noise cryptography which, however, were often developed with no reference to specific physical model. In those and some other proofs the sifted key $S$ is not \textit{expanded} by padded parity-check bits to form a codeword, rather it is \textit{contracted} to a shorter information bit sequence. Padding parity-check bits is then not applicable and B can compute the syndrome from the parity-check matrix of the linear ECC.  Such an ECC can be effectively employed by E to increase $\bar{p}_1(S)$ substantially to get her $\bar{p}_1(L)$, and it is not possible to pad a whole ECC with positive net key generation. In such case, $\rho^s_E$ must be first transformed to $\rho^l_E$ to be the input of the PAC, similar to the transformation of $\rho^l_E$ to $\rho^k_E$ as carried out in the security analysis of privacy amplification [5]. This density operator transformation has never been carried out [22]. In the following we give an example to show that disastrous breach of security may occur if the distinction of $S$ and $L$ in such situation is not maintained.

Consider a $(n,k)$ linear ECC with $2^k$ codewords of length $n=|S|$ such that $\bar{p}_1(S)\sim 2^{-k}$. Even when $k\ll n$, this $\bar{p}_1(S)$ could still be a number very small compared to 1, such as when $k$ is a significant fraction of $n$. Let $\rho^S_E$ be such that its range is partitioned into $2^k$ groups of orthogonal subspaces each of which is the range of $\rho^S_E$ for $2^{n-k}$ of the $s$ values. Let the openly known ECC assign each m-bit sequence in each group to just one sequence of the group in each of these orthogonal sub-ranges of $\rho^S_E$. This is a very good ECC indeed and the ECC can be linear from codeword assignment. Since the ECC structure is openly known, E can find $L$ exactly by measuring the projections into the orthogonal sub-ranges of the groups of $\rho^s_E$. On the other hand, the original $\rho^S_E$ indeed leads to a $\bar{p}_1(S)\sim 2^{-k}$ if the $\rho^s_E$ within one sub-range are very close in trace distance. The same idea clearly works classically also by replacing density operator with classical distribution. Thus, in this situation the $L$ is totally compromised. The idea of this construction generalizes in an obvious way to lesser security breach of various degrees. It seems applicable to quantum code protocols but not to entanglement purification protocols. In this paper, only BB84 type protocols are considered which have no entanglement or quantum code.

\section{Security of the Generated Key}
During A's quantum signal transmission E sets her probe and the protocol goes forward after checking. The final key $K$ is generated with corresponding ``prior probability'' $p(k)$ and probe state $\rho^k_E$ on each $k$. Let $\rho = \sum_k p(k)\ket{k}\bra{k}$ for $N$ orthonormal $\ket{k}$'s, $N=2^{|K|}$. Let $\rho_E = \sum_k p(k) \rho_E^k$, $\rho_{KE} = \sum_k p(k)\ket{k}\bra{k}\otimes \rho_E^k$. The trace distance criterion $d$ is defined to be \begin{equation}
d \equiv \frac{1}{2}\parallel\rho_{KE}-\rho_U\otimes\rho_E\parallel_1
\end{equation} where $\rho_U$ is $\rho$ with $p(k)=U(k)$ for the uniform random variable $U$ [23]. A key $K$ with $d\leq\epsilon$ is called ``$\epsilon$-secure'', as it has been forced by privacy amplification to be $\epsilon$-close to $U$. The major interpretation that has been given to $d\leq \epsilon$ amounts to saying perfect security is obtained with a probability $\geq 1-\epsilon$. In [4] it is explicitly stated ``The real and the ideal setting can be considered identical with probability at least $1-\epsilon$.'' In [24, 1] it is expressed with a different nuance with $\epsilon$ understood as ``maximum failure probability'' of the protocol  ``where 'failure' means that 'something went wrong', e.g., that an adversary might have gained some information on K''.

The justification of such erroneous interpretation of $d$ is derived from the interpretation of Lemma 1 in [4] that the variational distance $v(P,Q)$ between two distributions $P$ and $Q$ on the same sample space, the classical counterpart of trace distance, ``can be interpreted as the probability that two random experiments described by $P$ and $Q$, respectively, are different.'' That this interpretation cannot be true in any situation has been discussed in [6,7]. Here we give a simple example to bring out why.

Consider the distribution upon a measurement result with $P_i=\frac{1+2\epsilon}{N}$ for $i\in \overline{1-\frac{N}{2}}$ and $P_i=\frac{1-2\epsilon}{N}$ for $i\in \overline{(\frac{N}{2}+1)-N}$, so that $v(P,U)=\epsilon$. Then E ``gains information'' compared to the ideal case with probability 1/2, not $\epsilon$. This example clearly shows that variational distance is not the maximum probability that information is leaked. In fact, it is easy to see [6] that for $d > 0$, $K$ is \textit{not} uniform with probability 1 instead of being uniform with probability $1-d$. Thus, the operational meaning of $d$ has to be derived anew.

The operational security significance for $d$ can be derived from the classical properties of the variational distance between $K$ and the uniform distribution $U$. Upon a measurement result $y$ on her probe, E would derive a conditional distribution $p(k|y)$ on the possible $K$ values [2]. The $Y$-averaged $p_1(k|y)$ for the maximum of $p(k|y)$ given $y$ is the same as the $K$-averaged $\bar{p}_1(K)$ obtained from quantum detection theory [10]. We have 
\\*Theorem 1: \begin{equation}
\overline{p}_1(K) \leq \frac{1}{N} + \epsilon
\end{equation} which can be proved as follows. For each $y$, a $d(y)$ applies with average given by $d$. The maximum $p_1(y)$ is given by $d(y)+\frac{1}{N}$ [2,6]. Averaging over $Y$ gives (5) and we have also shown the bound may be obtained with equality.

Theorem 1 can be rewritten as a lower bound on $d$, \begin{equation}
d \geq \overline{p}_1(K) - \frac{1}{N}
\end{equation} that has a major implication on QKD security. The ECC and PAC are both known to E before she makes her probe measurement. Thus it is immediate from the many-to-one compression in both codes that \begin{equation}
p_1(S|y) \leq p_1(L|y) \leq p_1(K|y)
\end{equation} is true for each $y$. Indeed, it is typical that $\leq$ is $<$ with a relatively large gap. It follows from (6)-(7) that the $d$-level that can be obtained for $K$ satisfies $d(K) \geq \overline{p}_1(L)-2^{-|L|}$. Note that from (7), $\overline{p}_1(K)$ cannot be correctly bounded without explicit description of the ECC involved. E's PAC information is accounted for in the theory of [5] involving PAC averaging. On the other hand, the specific EEC information of E is \textit{not} accounted for in existing security proofs [22] unless the expanded parity-checks approach described in section 2 is carried out.

It also follows from (5) that $K$ is guaranteed to be near-uniform only when $d \lesssim\frac{1}{N}$, for example when $d =\frac{1}{N}$. In addition, Markov inequality [12] for the probability of a non-negative random variable $X$ taking large value \begin{equation}
P_r[X \geq \delta] \leq \frac{E[X]}{\delta}
\end{equation} needs to be used to convert it to an individual guarantee that the probability $X=p_1(K)$ is larger than an acceptable level must be small. Let us define, as in [1], the \textit{failure probability} $P_f$ to be the probability of failing to guarantee the non-occurrence of an unfavorable event, in this case that E identifies $K$ successfully. Note that this is a dramatic failure, and for security it must be guaranteed that E not be allowed to have an appreciable probability that it may happen. That is, one needs (8) to get an operational guarantee on $P_f$. Since only $\bar{d}$ as average over PAC is bounded [5] in the form $\bar{d} \leq \epsilon$, Markov inequality needs to be applied to get an individual PAC guarantee also. These two applications of Markov inequality are independent. Thus, we have from (8) for $d=1-\sigma$, $P_f \leq 1- (1-\sigma)(1-\frac{\epsilon}{\sigma})=\sigma+\frac{\epsilon}{\sigma}-\epsilon$. The best guarantee is obtained at $\sigma = \epsilon^{\frac{1}{2}}$ with resulting $P_f \sim 2\epsilon^{\frac{1}{2}}$ for $\epsilon \ll 1$. Similarly, for two applications of Markov inequality the total failure probability is minimized at $\sigma_1=\sigma_2=\epsilon^{\frac{1}{3}}$ with reality $P_f \sim 3\epsilon^{\frac{1}{3}}$. This shows that a cube root is needed on $\bar{d} \leq \epsilon$ to guarantee $p_1(k|y) \leq d^{\frac{1}{3}}+1/N$ for application to a specific PAC and E's observed $y$.

Note that the incorrect interpretation of $d$ as maximum failure probability would lead to a stronger statement than Theorem 1, because it asserts $p_1(K|y)=1/N$ with probability $\geq 1-\epsilon$ under $d \leq \epsilon$. More specifically, it implies with just one probability $\geq 1-\epsilon$ that any number of different uses of $K$ and subsets of $K$ all behave as if $K$ is uniform. On the other hand, the total probability of such uses would actually approach zero exponentially.

\section{Secure Key Rate}
We have seen above that one needs to get the exponent of $d$ equal to the bit length of the sequence under consideration to regard it as near-uniform, We also see that Markov inequality needs to be applied twice to get operational guarantee, with a resulting $\epsilon^{1/3}$ bound on the failure probability. Following (7)-(8), a $\bar{d}$ level on $K$ is bounded via a $\bar{p}_1(L)$ constraint to the PAC input as given by the Leftover Hash Lemma [25,26]. There is no need to consider $\epsilon$-smooth quantities other than $\epsilon\sim 0$ for increasing key rate by sacrificing security, due to the incorrect interpretation of $d$ and the need to generate near-uniform key with sufficiently small value of $d$. From [25,5] and abbreviating $|K|=n$, one may therefore set $\frac{[H_{min}-n]}{6}$ to be equal to $n$ to get a final $d$ level for $K$ with an exponent equal to $n$, making $K$ near-uniform from (5). This gives $n=\frac{H_{min}}{7}$ with $H_{min} >|S|\cdot[1-h(Q+\mu)]$ from (4)-(5) in [3], where $\mu$ is an $\epsilon$-dependent factor obtained for joint attack with an invalid classical counting argument. If we assume anyway this (4)-(5) in [3] gives an accurate estimate, for $\epsilon\sim 0$ it would push $h(Q+\mu)$ toward the value 1 and equ(4) in [3] does not then give a useful bound on $H_{min}$. Nevertheless, we proceed with $\mu=0$ for collective attacks so that the net key generation rate is obtained by subtracting $n(\frac{1}{r}-1)$ bits where $r$ is the rate of the actual linear ECC employed. Thus, a new key can be generated if \begin{equation}
h(Q)<8-\frac{7}{r}.
\end{equation}For the best possible asymptotic $r=1-h(Q)$, (9) shows net key can be obtained for $Q$ up to $\sim 1.5\%$. If one uses practically decodable ECC with reasonable error performance, for example the $(m,n)=(8160, 7159)$ QC-LDPC code in [15, p.502], no net key can be generated.

The secure key rate calculated in [3] has a very large associated $\bar{d}$ with $\epsilon^{1/3}> 10^{-5}$ even for very small key rate. One cannot exchange such key for the uniform secret key bits in $leak_{EC}$ that is needed for security proof, and our above analysis is the correct approach. In the NEC system [27] which is the only experimental QKD system with quantified security and which is a near complete protocol missing just message authentication, the ECC information to E is not accounted for. This is a possibly risky gap as we showed in section 2. Also, $Q > 5\%$ in that decoy-state system is too large for net key generation with the padded bits approach, if one recalls that the above $1.5\%$ $Q$ limit is for an ideal single-photon scheme under collective attacks. Indeed, a very negative key rate results when (1) is applied with $f=1$.

\section{Conclusion}
Security is a quantitative issue. There is unconditional (information-theoretic) security for conventional symmetric-key ciphers also [6]. We have shown that no concrete QKD protocol has been given that has a provable net secure key rate even when the necessary message authentication steps in a disturbance-information tradeoff QKD protocol are not integrated into the protocol and bits consumed not accounted for. Furthermore, there are other invalid classical counting steps in the security analysis. Many relevant physical system characteristics are not included such as system loss and detector electronic behavior. See also [28, 29]. Fundamental quantitative security against known-plaintext attacks in one-time pad use of the key is yet to be established. It does appear alternative approaches need to be explored for secure key generation.

\section{Acknowledgement}
This work was supported by the Air Force Office of Scientific Research. I would like to thank Greg Kanter for useful discussions.


\begin{thebibliography}{9}
\bibitem{ref16} There is much confusion in the QKD literature on A covering the ``syndrome'' with shared secret-bits, but the ECC syndrome is only known to B. It is the parity-check bits that are being covered. This is also distinct from the syndrome or coset announcement in the Shor-Preskill approach which effectively specifies a PAC. Further confusion is engendered from the use of entanglement purification protocols and BB84 type protocols of different variety.
\bibitem{ref17} It will be show elsewhere that the known proof on such optimality of collective attacks is not valid.
\bibitem{ref19} This is an ``in principle'' issue. If one can deal with exponential complexity it would be ``disruptive technology'' indeed.
\bibitem{ref23} The added freedom of having a state different from $\rho_E$ in (4) given in [3] is not applicable when $d$ is interpreted operationally in terms of E's success probability, such as that of Theorem 1 in this paper.
\end{thebibliography}
\end{document}